# An approximate analytical (structural) superposition in terms of two, or more, α-circuits of the same topology: Pt.1 – *description* of the superposition

*Emanuel Gluskin*

Galilean Sea Academic College, Holon Inst. of Technology, 58102, *and* Electrical Engineering Department of the Ben-Gurion University, Beer-Sheva 84105, Israel.
Email: gluskin@ee.bgu.ac.il.

**Abstract**: One-ports named "*f*-circuits", composed of similar conductors described by a *monotonic polynomial*, or *quasi-polynomial* (i.e. with positive but not necessarily integer, powers) characteristic $i = f(v)$ are studied, focusing on the algebraic map $f \to F$. Here $F(.)$ is the input conductivity characteristic; i.e., $i_{in} = F(v_{in})$ is the input current. The "power-law" "$\alpha$-circuit" introduced in [1], for which $f(v) \sim v^{\alpha}$, is an important particular case. By means of a generalization of a parallel connection, the *f*-circuits are constructed from the $\alpha$-circuits of the same topology, with different $\alpha$, so that the given topology is kept, and '*f*' is an additive function of the connection. We observe and consider an associated, *generally approximated, but, in all of the cases studied, always high-precision*, specific superposition. This superposition is in terms of $f \to F$, and it means that $F(.)$ of the connection is close to the sum of the input currents of the independent $\alpha$-circuits, all connected in parallel to the same source. In other words, $F(.)$ is well approximated by a linear combination of the same degrees of the independent variable as in $f(.)$, i.e. the map of the characteristics $f \to F$ is close to a linear one. This unexpected result is useful for understanding nonlinear algebraic circuits, and is missed in the classical theory.

   The cases of $f(v) = D_1 v + D_2 v^2$ and $f(v) = D_1 v + D_3 v^3$, are analyzed in examples. Special topologies when the superposition must be ideal, are also considered. In the second part [2] of the work the "circuit mechanism" that is responsible for the high precision of the superposition, in the most general case, will be explained.

## 1.   Introduction

### *1.1.*   *The circuit*

In its general form, the circuit under discussion is shown in Fig. 1. This is a 1-port of arbitrary structure composed of similar algebraic elements, here conductors. One can always define the topology so that each branch includes only one conductor *f*, though sometimes it is appropriate to speak about branches which include several series conductors.

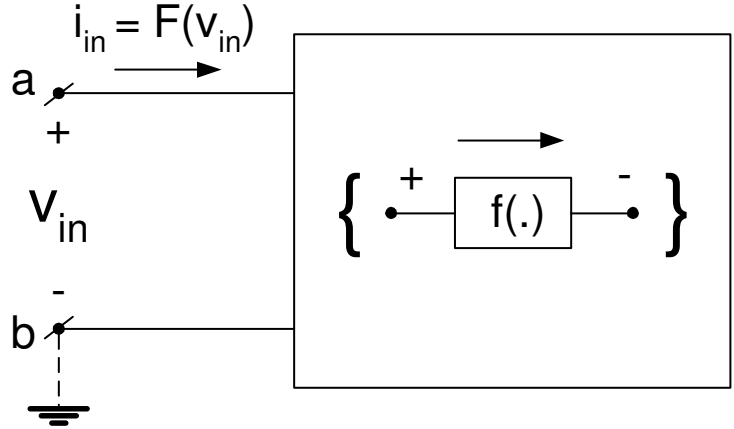

Fig.1: The 1-port (the "*f*-circuit") of a given topology, composed of similar conductors *f*(.). Magnetic and dielectric d.c. realizations are also possible. The most typical case below is when '*f*' is a polynomial having two terms, and we speak then about a "polynomial circuit". For the one-term (power-law) $f(v) \sim v^\alpha$, $\alpha > 0$; we speak about an $f_\alpha$-circuit, or "$\alpha$-circuit" (the latter circuits were introduced in [1]). The $\alpha$-circuits are the building blocks in our constructions.

Definition 1: We call such a 1-port an "*f*-circuit" and study for it the map $f \to F$, where *F* is the input current, $i_{in} = F(v_{in})$. □

Examples of "*f*-circuits" are the infinite homogeneous grids (1-ports), the nonlinear of [3-5], or the well-known linear one of [6], and some finite grid-cut type circuits [4,5,7].

In the description of the *f*-circuits, index *s* will label the branches, and index *k* the nodes. We shall write KCL at the input node, expressing the input current $i_{in} = F(v_{in})$ via the internal currents that are close to the input. For this we label the branches that enter node **a** by *s'*, and those entering node **b** by *s''*. In agreement with these notations, the *node* that is directly connected to **a** by a certain branch *s'* will be denoted by subscript $k_{s'}$, and the *node* that is directly connected to **b** by a certain branch *s''* will be denoted by subscript $k_{s''}$. If there is a conductor directly connecting **a** and **b**, then **a** belongs to the nodes $\{k_{s''}\}$, and **b** belongs to the nodes $\{k_{s'}\}$. Such a conductor simply adds $f(v_{in})$ to $F(v_{in})$, and the very possibility of this addition suggests some *analytical similarity* between $F(.)$ and $f(.)$ in a more general case too.

The input current $i_{in} = F(v_{in})$ can be expressed, by means of an input KCL equation, using the potentials of either only the nodes $k_{s'}$, or only $k_{s''}$. Since **b** is grounded, it is most appropriate to use the internal currents combined at **b**, and we shall prefer $k_{s''}$. Thus, the branch voltages $v_{s''}$ will be most commonly met below.

For a general case, the number of parallel conductors between a node $k_{s''}$ and node **b** will be denoted as $w_{s''}$, though below, as a rule, all $w_{s''}$ equal 1. Thus, generally, the input KCL equation is:

$$F(v_{in}) = \sum w_{s''} f(v_{s''}), \qquad (1)$$

where the dependence on $v_{in}$ comes via that of $v_{s''}$.

## 1.2. The case of the α-circuit

The characteristic $f(.)$ is always assumed to be monotonic, at least in the actual region of the independent variable, and such that $vf(v) > 0$, i.e. the elements are passive. This is essential ([1,8-11]) for the existence of a unique circuit solution. The case of $f(.)$ polynomial, or quasi-polynomial, i.e., including positive, not necessarily integer, powers of $\alpha$, is most important.

An analytically very useful case is the odd power-law characteristic

$$f(v) = D|v|^\alpha \text{sign}[v], \quad D, \alpha > 0,$$

for which the condition $vf(v) > 0$ is satisfied; $D|v|^{\alpha+1} > 0$. If $\alpha = 1$, then $f(v) = Dv$, i.e. the circuit is linear. Since the physical polarities of $\{v_s\}$ in the circuit digraph may be simply determined for the special linear case, we can always define the incident matrix of the circuit so that $v_s > 0, \forall s$. Thus, following [1], we simplify the above $f(.)$ to

$$f(v) = Dv^\alpha \qquad (2)$$

(i.e. $i_s = f(v_s) = Dv_s^\alpha$) and employ only this model for the case of the power-law $f(.)$.

Remark 1-1: Though the treatment below will be focused on the cases of integer $\alpha$, it is worth stressing that for a fractal $\alpha$ only the arithmetic (real, positive) value of any root is relevant in (2), and the positiveness of $v$ ensures that $f(v)$ in (2) is always real-valued (and positive). □

A suitable notation for (2) is $f_\alpha(.)$, and, following [1], we shall also name the $f_\alpha$-circuit the "$\alpha$-circuit". When speaking about applications for which only integer $\alpha$ are involved, $\alpha = p$, $p \in \mathbb{N}$, notation $f_p(.)$ ("$f_p(.)$-circuit") will be used.

The coefficient $D$ does not influence the nodal potentials of any $\alpha$-circuit, because it is contracted from the KCL equations that define $\{v_k\}$ by writing $\{i_s\}$ using (2). The independence of $v_k$ (and thus of the voltage drops $v_s$) from $D$ results in $i_s \sim D, \forall s$. Thus, for instance, by connecting to each of the circuit's elements one such in parallel, which means in (2) $D \to 2D$, we always just double the previously obtained $\{i_s\}$ and the input current $F(.)$.

One can thus make the whole circuit analysis for the case of $D = 1$, adding $D$ to the currents as a factor at the very end. Regarding the role of $D$, the map $f \to F$ is thus precisely linear.

From (2), $v = D^{-1/\alpha} i^{1/\alpha}$, and in the important asymptotic case of $\alpha \to \infty$ we find the conductors to be *voltage hardlimiters*, which makes the analysis of $\alpha$-circuits with large $\alpha$ very lucid. For instance, the infinite ladder shown in Fig. 5 below becomes, as $\alpha \to \infty$, a series connection of three similar voltage hardlimiters. However, in the most realistic cases, $\alpha$ has small values, usually 1, 2, 3, and the minimal value of $\alpha$, appearing below in equation (5) is, usually, 1.

Analysis of the $\alpha$-circuit may be not easy, but it is much easier than a precise analysis of an $f$-circuit with a polynomial $f(.)$, and it is possible ([2] justifies this insistence finally) that the $\alpha$-circuit will be a very useful tool.

It is shown in [1] that for any $\alpha$-circuit, $F(.)$ (or $F_\alpha(.)$) is analytically similar to $f_\alpha(.)$, i.e.

$$F(v_{in}) = F_\alpha(.) = D_\Sigma(\alpha) v_{in}^\alpha \qquad (3)$$

with some coefficient $D_\Sigma(\alpha)$ independent of $v_{in}$, which includes $D$ as a factor.

The function $\varphi(\alpha)$ defined by the equality

$$D_\Sigma(\alpha) = D\varphi(\alpha)$$

is the main characteristic of the map $f_\alpha \to F$, for the $\alpha$-circuits. Thus,

$$F(v_{in}) = D\varphi(\alpha)(\alpha) v_{in}^\alpha . \qquad (3a)$$

Another simple property of the $\alpha$-circuits is [1] that the nodal potentials satisfy the relation

$$v_k = d_k(\alpha) v_{in}, \quad \forall k, \qquad (4)$$

where all $d_k(\alpha)$ are independent of $v_{in}$. Since each of the branch voltage drops $v_s$ is a difference of some two of the $v_k$, (4) also yields

$$v_s \sim v_{in}, \quad \forall s . \qquad (4a)$$

Since the nodes close to the grounded terminal **b**

$$v_{s''} = v_{k_{s''}} - v_b = v_{k_{s''}}, \quad \forall s'',$$

using (1), (2) and (4), we obtain:

$$F(v_{in}) = \sum_{s''} w_{s''} f_\alpha(v_{s''}) = D \sum_s w_{s''} v_{s''}^\alpha = D(\sum_s w_{s''} d_{k_{s''}}^\alpha) v_{in}^\alpha ,$$

and, comparing with (3a), find

$$\varphi(\alpha) = \sum_{s''} w_{s''} d_{k_{s''}}^\alpha(\alpha) .$$

### 1.3. *The point of the research and the structure of the work*

The present research describes (and explains in the next part [2]) the *fact* (see also [3-5]) that when the map $f_\alpha \to F(.)$ is considered on a 1-port topology, and we can write $f_{\alpha_1} \to D_1 \varphi(\alpha_1)(\cdot)^{\alpha_1}$ and $f_{\alpha_2} \to D_2 \varphi(\alpha_2)(\cdot)^{\alpha_2}$, then, -- absolutely precisely for some topologies, and for many others (all those checked) with an unexpectedly high precision, -- we have for $f = f_{\alpha_1} + f_{\alpha_2}$ that

$$f_{\alpha_1} + f_{\alpha_2} \to D_1 \varphi(\alpha_1)(\cdot)^{\alpha_1} + D_2 \varphi(\alpha_2)(\cdot)^{\alpha_2} . \qquad (5)$$

That is, generally, *in terms of $f_\alpha$, $f \to F$ is, with a high precision, a linear map*.

Of course, this approximate linearity in terms of the polynomial *structure* has nothing in common with the usual *input-output* superposition of linear circuits, i.e. $F(v_{in})$ is, generally (i.e. if $\alpha_1$ and $\alpha_2$ are both not 1), *not* proportional to $v_{in}$. At the same time, the specific, *generally approximate*, "analytical", or "structural" superposition under study, is absolutely precise ("ideal") for linear circuits.

That the specific superposition may be ideal also for a very strongly nonlinear (and however complicated-structure) *f*-circuit becomes obvious if one takes the case of $\alpha_1 = \alpha_2$ when $f_{\alpha_1} + f_{\alpha_2} = 2f_{\alpha_1}$. This is equivalent to just doubling '*D*' in a separately taken $\alpha$-circuit, and thus to double $F(.)$, making it equal the sum of the input currents of the two separately taken similar $\alpha$-circuits.

The *approximate linear mapping of f(.) on the input conductivity function $F(.)$*, expressed by (5), is shown in [4-5] to be very helpful for calculation/estimation of $F(.)$ of the 1-ports for two-term $f(.)$, and the problem introduced in [7] of the comparison of the relative nonlinearity (curliness) of $F(.)$ with that of $f(.)$ may also be a field of application of the analytical superposition.

For a systematic treatment of the superposition a new type of circuit connection is introduced in Section 2.

Section 3 defines the "analytical superposition" in its general form, and proves an important general feature of the superposition.

Sections 4 and 5 describe specific cases when the situation as regards the superposition is very simple.

Sections 6 and Appendix A demonstrate for quasi-linear $f(.)$ the good precision of the "analytical superposition" using more usual examples. Since for any linear circuit the analytical superposition is ideal, and thus for a weakly nonlinear circuit, one would expect the analytical superposition to be highly precise without any special reasons, these examples demonstrate, in particular, that the error of the superposition is much smaller than the degree of the nonlinearity of the circuits. This very important point will be treated in detail in [2].

Section 7 collects some relevant data (partly acquired from [4,5]) in a table, showing the situation regarding precision more widely.

Section 9 explains the main steps that will be taken in [2].

Appendix B gives alternative *resistive formulation* of the $\alpha$-circuit, which in some cases can be more suitable than the conductive formulation.

## 2. The "*f*-connection" (or "$\alpha$-connection") of the 1-ports: $f(.)$ as an additive variable

Figure 2 illustrates the here basic concept of "*f*-connection". In this figure, two *f*-circuits (here $\alpha$-circuits, with the integer $\alpha = m$, $n$, and with $D = 1$) of the same topology are given in which *all* the pairs (two pairs are shown) of the respective nodes, including the input nodes **a** and **b**, are short-circuited.

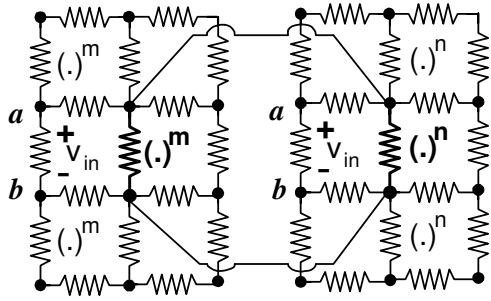

Fig. 2: The node-to-node "*f*-connection" of two circuits of the same topology, here one with $D_m \cdot (.)^m$ and another with $D_n \cdot (.)^n$. *All* the respective nodes are connected in pairs, as is shown for four nodes. Obviously, the resulting circuit is of the same topology, and $f(.) = f_1(.) + f_2(.)$, here $D_m(.)^m + D_n(.)^n$. Before the connection, the composing circuits are named "$f_m$-circuit" and "$f_n$-circuit", and after connecting, "$f_m^{cnct}$-circuit" and "$f_n^{cnct}$-circuit". Even though the $f_m^{cnct}$-circuit and the $f_n^{cnct}$-circuit are multi-ports, they may be easily defined. Their particular input currents "$F_m^{cnct}(.)$" and "$F_n^{cnct}(.)$, associated with connecting both of the '**a**'-nodes and both of the '**b**'-nodes to the terminals of the same voltage source, are also clearly defined, and for the whole connection $F(.) = F_m^{cnct}(.) + F_n^{cnct}(.)$, obviously.

<u>Definition 2</u>: The circuit (1-port) obtained by *short-circuiting all the respective nodes* of two, or more, *f*-circuits of the same topology, having the same $v_{in}$ at their inputs, will be named "*f-connection*".  □

<u>Definition 3</u>: The *f*-circuits *included in* (*composing*) the *f*-connection will be named "*f-connected circuits*", and denoted as $f^{cnct}$-circuits, or $f_p^{cnct}$-circuits, $p = 1,2, \ldots$ .  □

Each of the initially given $f_p$-circuits (1-ports) generally becomes a multi-port in the *f*-connection, and in this sense this generalization of the usual parallel connection is "destroying". Thus, we should have some firm rules for approaching $f_p^{cnct}$-circuits. However, these rules are very simple.

For each of the $f_p^{cnct}$-circuits, we always take its input current at the same port that remains directly connected to the same voltage source, and only this input current interests us finally. It is obvious which physical elements belong to any certain $f_p^{cnct}$-circuit; all these elements have the same characteristic $f(.)$ given for this $f_p$-circuit before the connection. Thus, the concept of the "$f^{cnct}$-circuit" is absolutely clear.

<u>Definition 4</u>: The input current of the $f_p^{cnct}$-circuit will be denoted as $F_p^{cnct}(.)$, $p = 1,2, \ldots, P$.  □

It is obvious that $F(.) = (\Sigma F_p^{cnct})(.)$. Thus, for instance, for the connection of Fig. 2, we have $F(.) = F_m^{cnct}(.) + F_n^{cnct}(.)$.

The analytical problematicity here is, of course, that contrary to the usual parallel connection, for the *f*-connection $\{F_p^{cnct}(.)\}$ are, generally, mutually dependent.

Since all the respective-branch elements become connected in parallel, the analytical meaning of the "*f*-connection" of $P$ *f*-circuits is simply that in the given topology we set

$$f(.) = f_1(.) + f_2(.) + \ldots + f_P(.) ,$$

i.e. mathematically, *f*(.) *is an additive scalar function on the given topology*.

As against the simplicity of this formulation, *the problem of the change in F(.) when f(.) is changed* is, generally, a very difficult one, and the study of this problem via the *f*- connection of the proper $\alpha$-circuits is methodologically justified.

When a polynomial *f*-circuit *is given*, e.g. with $f(.) = D_m(.)^m + D_n(.)^n$, we can *interpret it* as an *f*-connection, with the composing $\alpha$-circuits being uniquely defined. Thus, since the polynomial $D_m(.)^m + D_n(.)^n$ cannot be identically replaced using any other degrees, but $(.)^m$ and $(.)^n$, if $f(.) = D_m(.)^m + D_n(.)^n$ is given for the topology in Fig. 2, we can convert the procedure, interpreting the given final circuit by means of the '*f*-connection' of the two "wings" with $f_m(.)$ and $f_n(.)$.

Such an interpretation of the polynomial circuit, *intended to help in the analysis of its F(.)*, is named the "$\alpha$-test". Actually, we replace *f*(.) by $(.)^\alpha$ and calculate $D_\Sigma(\alpha)$, or $\varphi(\alpha)$, to be used as explained in the next section.

When dealing with the "$\alpha$-test", the term "$\alpha$-connection" will be sometimes used instead of the term "*f*-connection".

While speaking below, in general, about $\alpha$-circuits with any positive $\alpha$, we shall proceed with integer degrees, which seem be most practical.

### 3. The approximate analytical superposition

Consider the precise $F(v_{in})$ of an *f*-connection with a polynomial '*f*'. In the following formulae, index *p* both denotes the degrees and labels the circuits. Thus, $f_p = D_p(.)^p$, and for *f*(.) and *F*(.) of the whole connection we have, respectively,

$$f(\cdot) = (\sum_p f_p)(.) = \sum_p D_p(\cdot)^p, \qquad (6)$$

and

$$F(v_{in}) = \sum_p F_p^{cnct}(v_{in}). \qquad (7)$$

We consider the approximation of $F(v_{in})$ by the function (in each its term, (3a) is used)

$$G(v_{in}) \equiv \sum_p F_p(v_{in}) = \sum_p D_{\Sigma_p}(p) v_{in}^p = \sum_p \varphi(p) D_p v_{in}^p \qquad (8)$$

which is the sum of the *independent* (*before the f-connection*) input currents of the involved $\alpha$-circuits, having $v_{in}$ at their inputs, i.e. simply connected in parallel.

<u>Definition 5</u>: Approximation (replacement) of the precise *F*(.) by *G*(.) will be named "*approximate analytical (or structural) superposition*". For brevity of writing, we shall often use one word, "*superposition*". □

In all the cases, the studied function *G*(.) appears to be unexpectedly close to *F*(.), and the possibility of having such a simple approximation for *F*(.) as *G*(.), even if only as a "first approximation", is a theoretically interesting point, especially when considered against the exceptional difficulty of determining (even for a circuit having a relatively simple structure [4,5]) the precise *F*(.).

Remark 3-1:   In [3-5], the power series expansion

$$F(v_{in}) = \sum_p b_p v_{in}^p , \qquad (9)$$

with some coefficients $\{b_p\}$ (that are very bulky and difficult to obtain), is used instead of the general form (7), and this series is then truncated to have the same number of terms as in *f*(.).  Though such a series expansion of the precise *F*(.) is possible only for some limited $v_{in}$, it provides us with the important possibility of observing the superposition in a continuous range of $v_{in}$.  Unfortunately, such series expansions are never really helpful in seeing the *reasons* [2] for the superposition, and we shall be focused only on the qualitative theory that reveals these reasons satisfactorily.  □

Definition 6:   The *relative error of the superposition* is defined as the relative *nonnegative* difference

$$\eta = \left|\frac{G-F}{F}\right| = \frac{|G-F|}{F} = \frac{|F-G|}{F} . \qquad (10)$$

In numerical examples, this error will be often presented in %.  □

Since the *input power* of the *f*-connection (the "*F*-circuit") is

$$P_F = P_F(v_{in}) = v_{in} F(v_{in})$$

and that of the usual parallel connection (the "*G*-circuit"):

$$P_G = P_G(v_{in}) = v_{in} G(v_{in}),$$

we can rewrite (10) as

$$\eta = \frac{|P_F - P_G|}{P_F} . \qquad (10a)$$

Using the input powers instead of the input currents allows one to compare the "*F*-circuit" with the "*G*-circuit" in terms of (1), i.e. *not in terms of only the nodes {s"} which are close to the input, but in terms of all of the circuit's nodes {s}*. The latter makes it possible to apply [2] the power (in particular, Tellegen's) theorems to the analysis of the *f*-connection.

It is obvious that similar positive term(s) in *F*(.) and *G*(.) decrease *F*(.) - *G*(.) and $\eta$.

Sometimes it is suitable to compare only the nonlinear parts of *F* and *G*. The relative distinction between the nonlinear parts is always somewhat larger than $\eta$.

### *3.1. The statement about the identity of the linear terms of F(.) and G(.)*

The following feature of *G(.)* or *F(.)* is observed in all examples, and is absolutely general. It can be derived, in principle, from the implicit function theorem [12] that is relevant to any *precise* solution of the nonlinear circuit equations that define *F(.)*, but the circuit nature of the problem allows us to give a simple and more direct argument explaining the meaning of the following limit. While in the usual treatments of the implicit function theorem [12], the limiting case usually is the linear one, the characteristics here need not be quasilinear, i.e., $\min\{\alpha_p\}$ need not be 1.

Statement 1:

$$\lim_{x \to 0} \frac{F(x)}{G(x)} = 1 \qquad (11)$$

*for any topology. That is, the first term in the power expansion of G(.) coincides with the first term in F(.).* □

Since the proof separately considers the *f*-connection that defines *F(.)*, and the usual parallel connection that defines *G(.)*, it is also possible to write (11) as the ratio of finite nonzero limits:

$$\lim_{x \to 0} \frac{F(x)/x^{\alpha_1}}{G(x)/x^{\alpha_1}} = \frac{\lim_{x \to 0}[F(x)/x^{\alpha_1}]}{\lim_{x \to 0}[G(x)/x^{\alpha_1}]} = 1 \, ,$$

where $\alpha_1$ is the *minimal* value of the degrees involved. Additionally, we consider only one of the other degrees, denoted as $\alpha_2$. The latter limitation obviously does not limit the generality of the proof, because we always have a *finite number* of degrees in such a problem.

Proof: For the *parallel connection* that defines *G(.)*, we use (8) and see that in

$$G(v_{in}) = D_{\Sigma_1}(\alpha_1) v_{in}^{a_1} + D_{\Sigma_2}(\alpha_2) v_{in}^{a_2}$$

the ratio of the second term to the first one is

$$\frac{D_{\Sigma_2}(\alpha_2)}{D_{\Sigma_1}(\alpha_1)} v_{in}^{a_2 - a_1}$$

for which, since $\alpha_2 > \alpha_1$,

$$\lim_{v_{in} \to 0} \frac{D_{\Sigma_2}(\alpha_2)}{D_{\Sigma_1}(\alpha_1)} v_{in}^{a_2 - a_1} = 0.$$

That is, as $v_{in} \to 0$ the parallel connection becomes equivalent to only the realization of the digraph with $\alpha = \alpha_1$, $G(v_{in}) \sim D_{\Sigma_1}(\alpha_1) v_{in}^{a_1}$.

Turning now to the *f-connection* that defines $F(.)$, we have *in the branches* that

$$i_s = D_1 v_s^{a_1} + D_2 v_s^{a_2}$$

and the ratio of the second term to the first one is (using (4a))

$$\frac{D_2}{D_1} v_s^{a_2-a_1} \sim v_{in}^{a_2-a_1} \to 0, \text{ as } v_{in} \to 0.$$

That is, only the element with $f(.) \sim (.)^{\alpha_1}$ has influence.

It appears that both the *f-connection* and the parallel connection, which define, respectively, $F(.)$ and $G(.)$, become as $v_{in} \to 0$ *the same circuit* with $f(.) \sim (.)^{\alpha_1}$. Thus, $F(.)$ and $G(.)$ must also become the same as $v_{in} \to 0$, and since for $v_{in} \to 0$ the main terms in the series for $F(.)$ and $G(.)$ are the first terms, these terms must be identical. □

Since in $F - G$ the equal ($\sim (.)^{\alpha_1}$) terms of $F$ and $G$ are cancelled, this first term in $F(.)$ influences $\eta = |F - G|/F$ only via the denominator $F$, increasing this denominator and thus decreasing $\eta$. For a linear circuit, $F = G$ and $\eta = 0$, obviously.

Below, Statement 1 is illustrated by the example of Section 6 and the example of Appendix A.

## 3.2 The role of separated parallel branches as regards precision of the superposition (*a possibility to decrease $\eta$*)

We can ensure that $F(.)$ and $G(.)$ include some similar positive terms, by means of connecting some branches, each including only series elements, in parallel to **ab**. This necessarily improves the superposition. For instance, in the circuit shown in Fig. 3 the two first branches introduce the same terms in $F(.)$ and $G(.)$, increasing $\eta$.

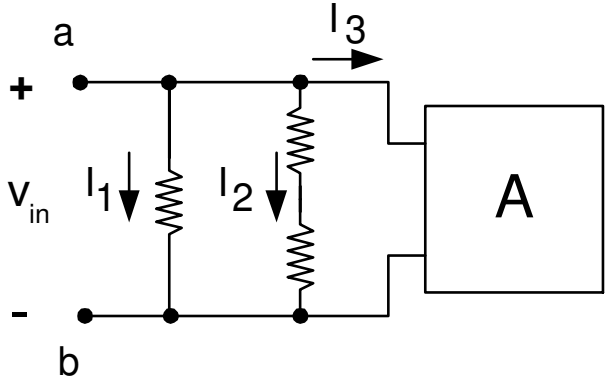

Fig. 3: A circuit including parallel branches that improve the analytical superposition. This is because these branches appear as independent 1-ports for which the superposition is ideal.

Here, $F(.) = i_1 + i_2 + F_3$ ($F_3(.) = i_3$) and $G(.) = i_1 + i_2 + G_3$, with the same $i_1$ and $i_2$, and all terms are positive.

Regarding the definition of $G_3$ in the latter equation, we note that in the parallel connection of any two *f*-realizations of such a circuit, the involved subcircuits "**A**" are also seen to be connected *in parallel*. Parallel input branches always leave the rest of the circuit as a 1-port, and they just add into $F(.)$ and $G(.)$ the same terms related to "**A**".

The superposition for the whole circuit is thus necessarily increased because the denominator of the fraction $\eta = |F - G|/F$ is increased.

The infinite square grids, or their cuts (e.g. [3,4]), as 1-ports, usually include such a "central" (**ab**) element. This element increases the (input) nonlinearity of the whole circuit, while the added common term decreases the error of the analytical superposition. See the example in Section 6.

## 4. The case when the analytical superposition is absolutely precise ("ideal")

There are, in total, three reasons, -- having, however, very different degrees of generality, -- for the high precision of the analytical superposition. Firstly, there are some *special cases*, associated with requirements related to the circuit's topology, where the superposition is *precise*, i.e. $G(.) \equiv F(.)$. Secondly (Section 5), when only *large values of $\alpha$* are involved, a very high precision of the superposition takes place for any topology. The last and, really, general reason for the precision, relevant to any topology of the 1-port, and any values of $\alpha$ (or $p$) is [2] that the differences $\{F_p^{cnct}(.) - F_p(.)\}$ have different signs for different $p$. That is, *for one of the circuits involved in the f-connection, the input current is increased, and for the other decreased*, and thus the obtained $F(.)$ differs from $F_{p1}(.) + F_{p2}(.)$ weakly. This very interesting fact is explained in [2] by an analysis of the dependences $v_k(\alpha)$. Let us start, however, from the simplest case.

If in the $\alpha$-circuit, the nodal voltages $\{v_k\}$ are absolutely independent of $\alpha$, then for all the $\alpha_p$-realizations of the same digraph, the potentials of the respective nodes are the same. Since, then, "*f*-connecting" means connecting points with the same potentials, this connecting changes nothing. We thus obtain a circuit that is equivalent to the usual parallel connection of the given 1-ports, i.e. $G(.) \equiv F(.)$ *precisely*, and the analytical superposition is ideal.

For this case, defined solely by the topology, the nodal voltages are independent not only of $\alpha$ for $f = (.)^\alpha$, but of any $f(.)$ ascribed to the elements. Thus, for the circuit of Fig. 4, with the common ground at node 'b', $v_c = v_e = (2/3)v_{in}$ and $v_d = v_f = (1/3)v_{in}$, independently of $f(.)$. This becomes clear if we delete the conductors *c-e*, and *d-f*, obtaining two similar voltage dividers, *a-c-d-b* and *a-e-f-b*, i.e. equal voltages in 'c' and 'e', and in 'd' and 'f'.

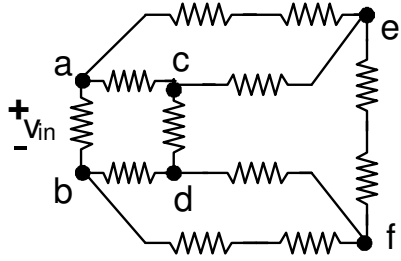

<u>Fig. 4</u>: A circuit for which the analytic superposition gives a precise result, regardless of *f*(.).

The *f*-connection of the *f*-circuits of this topology will not change each of the circuits and currents and will be equivalent to the usual parallel connection.

Using in this example potential $v_c$, or $v_d$, and parameters 'd' defined by (4) we obtain

$$\varphi(\alpha) \;=\; 1 + 2(1-d_c)^\alpha \;=\; 1 + 2d_d^\alpha \;=\; 1 + 2(1/3)^\alpha.$$

Using this $\varphi(\alpha)$ for two such *f*-connected $\alpha$-circuits, one with $\alpha = m$, and another with $\alpha = n$, we obtain (8) as

$$G(v_{in}) \;=\; F_m(v_{in}) + F_n(v_{in}) \;=\; D_m\,[1 + (1/3)^m]\,v_{in}^{\,m} \;+\; D_n\,[1 + (1/3)^n]\,v_{in}^{\,n}$$

which is also *the precise F*($v_{in}$), with the terms rearranged. Indeed,

$$F(v_{in}) \;=\; [D_m\,v_{in}^{\,m} + D_n\,v_{in}^{\,n}] \;+\; [D_m\,(1/3)^m\,v_{in}^{\,m} + D_n\,(1/3)^n\,v_{in}^{\,n}]$$

where the first bracketed term relates to the **ab**-conductor.

Circuits with nodal voltages completely independent of $\alpha$ are exceptional, but for a circuit of a close structure, the superposition may be quite precise.

A simple case of the ideal superposition also is the case of $\alpha$-connection with all $\alpha_p$ equal (e.g., Fig. 2, for $m = n$). This changes only '*D*' in an $\alpha$-circuit while not influencing $\{v_k\}$, for any topology.

Comparing Fig. 3 with Fig. 4, we note that in the circuit of Fig. 3 *only some* of its $v_k$ (related to the nodes of the separated parallel branches, between any two sequential elements) are unchanged with *f*-connection, while in Fig. 4 all the $v_k$ are unchanged.

## 5. The *asymptotically* precise analytical superposition (the case of large $\alpha_p$)

As $\alpha \to \infty$, the elements of the $\alpha$-circuit become voltage hardlimiters, as is clear from the inversion of (2), $v \sim i^{1/\alpha}$. Each such hardlimiter is "blocking" the remainder of the circuit, and thus the transfer to hardlimiters finally leaves (as relevant to *F*(.)) only some series branches, all connected in parallel to the port. Then [3,1,4,5] the nodal voltages quickly tend to certain limits that are solely defined by circuit topology, and the situation becomes close to that discussed in Section 4. Already $\alpha = 3$ is a large value here, and, e.g., for $f(.) = D_3(.)^3 + D_5(.)^5$, $F(.)$ may be well obtained as the analytical superposition, for any topology.

<u>Definition 7</u>:   The *asymptotically precise, as* $\alpha_p \to \infty$, $\forall p$, approximation of $F(.)$ by $G(.)$ is named "*asymptotic analytical superposition*", or, briefly, "*asymptotic superposition*". □

It is easy to illustrate the "*asymptotic superposition*", using any of the circuit examples given here or in [4,5]. For instance, the circuits in Figs 3, 4 and A1 become only *one* input hardlimiter, and the infinite ladder shown in Fig. 5 below becomes *three* series similar hardlimiters.

## 6. An example: the infinite nonlinear ladder

We now turn to a more regular example, demonstrating the typical precision of the analytical superposition, obtained not because of a specific topology, or large values of $\alpha$. In this example, as well as the example of Appendix A, $f(.)$ is quasi-linear, i.e. $\min\{\alpha\} = 1$.

Consider, following [4], the infinite nonlinear ladder, shown in Fig. 5, composed of the conductors

$$f(.) = D_1(.) + D_2(.)^2 .$$

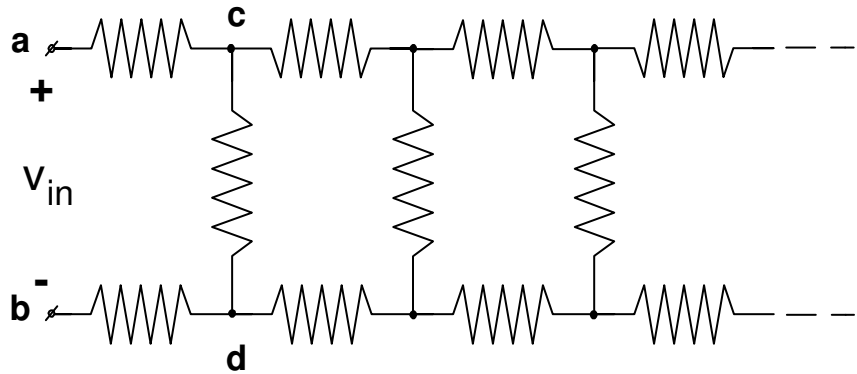

<u>Fig. 5</u>:  The infinite ladder of the nonlinear conductors $f(.) = D_1(.) + D_2(.)^2$ . In [4], we first calculate $F(.)$ precisely, which proved to be very difficult, and then use the "$\alpha$-test", i.e. interpret the circuit as the *f*-connection of the $\alpha$-circuits, using $G(.)$ for a much easier estimation of $F(.)$. Here, we first consider here the ladder as it is. Then, we add a conductor directly between the input nodes. Finally, we consider such a *two-directional* ladder, with the "central" conductor.

A solution based on precise circuit equations leads ([4] for very bulky details) to

$$F(.) = \frac{D_1}{1+\sqrt{3}}(.) + 0.1196 D_2(.)^2 \qquad (12)$$

The series expansion involved in the derivation of this two-term expression requires

$$v_{in} < 0.574 \frac{D_1}{D_2} . \qquad (13)$$

It is also found in [4], in terms of the "$\alpha$-test", that for the given topology, introducing

$$\lambda = v_{in}/v_{cd},$$

we have

$$\varphi(\alpha) = \left(\frac{\lambda-1}{2\lambda}\right)^\alpha .$$

<u>Remark 6-1</u>:  $\lambda = \lambda(\alpha)$ is [4] the largest (>1) positive root of the equation:

$$(\lambda^\alpha - 1)(\lambda-1)^\alpha = (2\lambda)^\alpha ,$$

from which $\lambda \to 3$ as $\alpha \to \infty$, i.e. $v_{cd} \to v_{in}/3$. This voltage division agrees with the fact that the circuit becomes three series hardlimiters as $\alpha \to \infty$. □

<u>Remark 6-2</u>:  $\lambda(\alpha)$, and thus $v_{cd}(\alpha)$, are ([4] for details) are monotonic functions. That any

$$d_k(\alpha) = v_k(\alpha)/v_{in}$$

is a monotonic function of $\alpha$ is a general rule for the $\alpha$-circuits. One can check this for any example given here (see especially Section A3 in Appendix A) and in [4], but the general reason is very simple; there is no other non-dimensional parameter given in the formulation of the problem, with which $\alpha$ can be compared, and in which $d_k(\alpha)$ can have an extremum. The argument re transfer to hardlimiters as $\alpha \to \infty$ also supports this. □

Using the above expression for $\varphi(\alpha)$, $G(.)$ is constructed according to (8):

$$G(.) = \frac{D_1}{1+\sqrt{3}}(.) + 0.115146 D_2(.)^2 . \qquad (14)$$

While the linear term is (as it must be, according to Statement 1) the same as in (12), the relative error in the coefficient in the quadratic term between (14) and (12) is 3.7%.

In agreement with the argument of Section 3.2, this relatively large error is strongly decreased when we add to the ladder the nonlinear conductor that directly connects nodes **a** and **b**. Then, as is easy to see, $f(.) = D_1(.) + D_2(.)^2$ is added to both (12) and (14), and denoting the new '$F$' and '$G$', as $F^+$ and $G^+$, we have:

$$F^+(.) = D_1(1+\frac{1}{1+\sqrt{3}})(.) + 1.1196 D_2(.)^2$$

and

$$G^+(.) = D_1(1+\frac{1}{1+\sqrt{3}})(.) + 1.115146 D_2(.)^2 .$$

The relative difference between the nonlinear parts of $F^+$ and $G^+$ is now $(0.1196/1.1196) \cdot 3.7\% = 0.4\%$, which is a very small value. The usual error for circuits including the conductor directly connecting **a** and **b** is 0.7-0.8%.

A *two-directional* (infinite on both sides) version of a ladder with such a "central" conductor is considered in [4], and the error in the analytical superposition is found there to have the intermediate value of 0.6%.

For reference purposes, let us also complete the data related to the infinite ladder of Fig. 5 by the calculated value $\lambda = 3.024688$, related to $\alpha = 3$. This gives $\varphi(3) = 0.03749$, to be used for obtaining $G(.)$ for the ladder with the *odd* characteristic $f(.) = D_1(.) + D_3(.)^3$ which may be a more practical model for a realistic $f(.)$, especially if one considers d.c. saturated magnetic of ferroelectric structures.

### *6.1. The precision of the superposition against the degree of the nonlinearity, for the two ladders* (*why do we consider the error of the superposition to be a small one?*)

Let us compare the precision of the superposition with the degree of nonlinearity of the circuit, in the versions of the ladder without and with the element that directly connects **a** and **b**. Substituting the upper limit for $v_{in}$, given by (13), into the ratio of the nonlinear and linear terms in $F$ (consider (12) with the argument $v_{in}$) i.e. into

$$0.1196(1+\sqrt{3})\frac{D_2}{D_1}v_{in},$$

we obtain, $0.1196(1+\sqrt{3})0.574 = 0.188$, this to be compared with the error in the superposition, 0.037.

In the second case, working with $F^+$ we substitute the upper limit for $v_{in}$ into

$$\frac{1.1196}{1+(1+\sqrt{3})^{-1}}\frac{D_2}{D_1}v_{in},$$

and obtain the relative nonlinearity as $(1.1196/1.366025)0.574 = 0.47$, this to be compared with the respective very small error of the superposition of 0.004.

*Thus, in each case, the error of the superposition is much smaller than the degree of the nonlinearity of the circuit.*

These examples, that use the results of the analysis of [4], relate to a continuous range of $v_{in}$. Appendix A presents a purely numerical example employing a certain $v_{in}$. The latter example is given in full detail, and though a little tedious, well shows what should not be missed in such analysis, and it is suggested that the Reader study it too.

### 7. Some collected data

Let us observe the precision of the analytical superposition in some "normal" cases (i.e. not the exceptional ones of Sections 4 and 5), collected in Table 1. The results relate to compositions with $\{\alpha\}$ given as $\{1,2\}$ and $\{1,3\}$.

| The circuit | "Degree of nonlinearity" | The relative distinction given is only **between the nonlinear terms**; the linear terms are identical. ($\eta$ is even smaller in all examples) |
|---|---|---|
| Fig. A1 of the present work (Appendix A).  $\{\alpha\} = \{1,3\}$ | 0.754 | **0.0046** |
| Infinite ladder (Fig. 5 of the present work). $\{\alpha\} = \{1,2\}$. | 0.188 | **0.037** |
| The infinite ladder (Fig. 5 with the added conductor that directly connects the input nodes **a** and **b**.) $\{\alpha\} = \{1,2\}$. | 0.47 | **0.004** |
| The two-directional cut of the two-directional nonlinear ladder given in Fig. 2 in [4]. $\{\alpha\} = \{1,2\}$. | Up to 0.5468 | **0.006** |
| The infinite 2D grid studied in [5]. $\{\alpha\} = \{1,2\}$. | 0.12 | **0.0083** |

Table 1:  Some relative errors, typical for the whole research, in the "analytical superposition" from some examples taken from the present work and [4,5].  The *nodal voltages* in all the cases are changed significantly (in the range of 5%-12%), and despite the also significant nonlinearity of the circuits, the error in the "superposition" is very small.

   Table 1 shows, for the low degrees of $\alpha$, that when the nodal voltages are significantly changed by the *f*-connection $G(.)$, compared with $F(.)$, the typical error is only about 1%.  Thus, the special conditions of Sections 4 and 5 are *not* necessary for the approximation of $F(.)$ by $G(.)$ to be a good one.

## 8.  A remark on many-ports

Both the physical *f*-connecting of the respective nodes, and the simple additive role of '*f*' on the analytical side, make it possible to perform an *f*-connection for any *multiports* of similar topology.  Then the nodal potentials become functions of all the input voltage sources, and for nonlinear circuits the situation becomes extremely difficult to analyze.  Consider, however, such a circuit as, e.g., a dense grid, having highly separated inputs connected to sources of similar intensity.  Then, in the vicinity of an (each) input, the branch currents (and thus also the input current of this input) will be mainly influenced by the close source, and we can approach a local part ("cut") of the circuit, around this input, as a 1-port.  In this case, we can obtain analytical superposition at each such 1-port. We can even allow $\alpha_1$ and $\alpha_2$ in $f(.) = D_1(.)^{\alpha_1} + D_2(.)^{\alpha_2}$ to be "slowly" changed as a function of the spatial coordinates defined in some way on the grid-type circuit, having the local respective analytical superposition near each port.

An analysis of the possibility of seeing the inputs of a multi-port as separated, must employ, however, the theory of distributed systems.

## 9. On the circuit mechanism responsible for the good precision of the analytical superposition for arbitrary topology of the f-circuit (brief plan of [2])

The main statements of the work [2] in which the "*m,n*-circuit" with $f(.) = D_m(.)^m + D_n(.)^n$ is considered, are listed below, each with brief comments.

**a** *For any k, the values of the nodal voltages $v_k$ of the "m,n-circuit" are between the respective values for the m-circuit and the n-circuit.*

Section A.3 of Appendix A here is a useful introduction to this point.

**b** *Thus, in view of (7), one of the currents $F^{cnct}$ is larger than the respective F, and another is lower, which is the mechanism of the "stabilization" of F, i.e. of keeping the relative error of the superposition $\eta$ small.*

We use $\alpha$-test, and consider that for the $\alpha$-circuit, monotonic dependence of all $v_{s''}(\alpha)$ may be different for different $s''$, i.e. some of $v_{s''}(\alpha)$ may increase and the other decrease with increases in $\alpha$, does not contradict statement **a**. We just have to separately consider these two subsets of the nodes, or branches, obtaining for the respective parts of $F(.)$ similar conclusions.

**c** *The ease of calculation of the $\alpha$-realization of the given 1-port topology allows one to present some bounds for $\eta$, using parameters $v_k(m)$ and $v_k(n)$ obtained for the separate m-circuit and n-circuit.*

This use of the $\alpha$-realization is seen in the scope of the general tendency/principle of the whole research to use that the $\alpha$-realization is much more simply calculated than any polynomial realization. We thus even assume that a *complete* solution (calculation) of the $\alpha$-realization for the relevant values $m$ and $n$ of $\alpha$ is relatively easy, and all $v_k(m)$ and $v_k(n)$ (or all $v_s(m)$ and $v_s(n)$) may be regarded as known. Under this assumption, [2] derives, in particular (here $D_m = D_n = 1$),

$$|F - G| < \frac{1}{v_{in}} \left( \sum_{\{s\}_1} [v_s^{n+1}(n) - v_s^{n+1}(m)] + \sum_{\{s\}_2} [v_s^{m+1}(m) - v_s^{m+1}(n)] \right)$$

where $\{s\}_1$ labels the voltage drops on the elements for which $v_s(n) > v_s(m)$, and $\{s\}_2$ labels the voltage drops on the elements for which $v_s(m) > v_s(n)$.

## 10. Conclusions and final remarks

We have considered algebraic 1-ports, composed of similar elements with a characteristic $f(.)$ including several (in the concrete examples 2) degrees. Such circuits not only generalize the single-degree ($f(.) \sim (.)^{\alpha}$) "$\alpha$-circuits" of [1], but also show their importance, via an analysis of a specific circuit composition, named "*f*-connection", which leads to the polynomial "*f*-circuits".

As the main point, the simplification in the estimation of $F(.)$ of the *f*-circuit, provided by the use of the approximate superposition, is very significant. While direct calculation of $F(v_{in})$ even only for a certain $v_{in}$, may be very difficult, the "$\alpha$-test" relatively easily gives the approximation $G(.)$ for $F(.)$, for every $v_{in}$.

The *f*-connection is "bad" as it belongs to the class of generally "destroying" connections, i.e. the type of the circuit to be connected is changed with the connection; being previously 1-ports, the composing circuits generally (besides the special cases of Section 4) become here multi-ports. "Destroying" connections are rarely mentioned in circuit theory, and when mentioned, it is mainly in the sense of the necessity of their prevention. (Thus, for instance, in the theory of 2-ports, the known Brune [13] tests are needed to prevent some "destroying" series and parallel connections of the 2-ports, *which change these 2-ports*.)

The similarity of the topologies (digraphs) of all of the circuits involved in the *f*-connection makes these circuits an interesting example of the application of Tellegen's theorem [2].

Work [7] considers some *relative smoothing of the nonlinearity of $F(.)$ with respect to that of $f(.)$* for *f*-circuits with polynomial characteristics. This may be relevant to lumped-circuit modeling of a resistive medium when the measurement of the conductive properties may be done only via input. One sees that the analytical superposition (or $\alpha$-test) could be effectively applied to such a problem by studying the curliness of $G(.)$ instead of that of $F(.)$, thus expressing the results in terms of $\varphi(.)$, obtained in the $\alpha$-test. Work [7] shows that $F(.)$ may be significantly less nonlinear than $f(.)$, and one performing the outer measurements of the conductivity (or, e.g., a d.c. magnetic characteristic, for the proper physical situation) has to know that.

Main proofs will be presented in [2], and we give in [2] alternative representations of the relative error of the superposition, one in terms of $\{v_{s''}\}$ and one in terms of $\{v_s\}$, using, in the latter case, *all the* nodal potentials of the circuit, or *all* branches' voltage drops.

The interesting *f*-connection and the feature of approximate analytical superposition for homogeneous resistive 1-ports are unjustly missed in the classical theory of resistive circuits.

## **Appendix A:** An example (full derivations) of direct investigations of the analytical superposition in a regular circuit, for a certain $v_{in}$

As an additional useful circuit example, let us compare $G(.)$ with precise $F(.)$, for the simple topology of Fig. A1, already employed in [1]. This time, we shall use a certain value of $v_{in}$, such that the nonlinearity is strongly revealed. The simplification of dealing with a numerical $v_{in}$, is justified by the reasons mentioned in Remark 3-1.

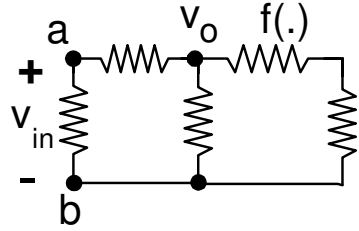

Fig. A1: The circuit for the analysis of the analytical superposition. First we take $f(.) = (.)^\alpha$ (the "$\alpha$-test") and then $f(.) = (.) + (.)^3$.

For writing $G(.)$, we use $\varphi(\alpha)$ found in [1] from the $\alpha$-version of the circuit ($f = (.)^\alpha$).

$$D_\Sigma(\alpha) = \varphi(\alpha) = 1 + \frac{1 + 1/2^\alpha}{\left(1 + (1 + 1/2^\alpha)^{1/\alpha}\right)\alpha} \quad . \tag{A1}$$

Choosing then the quasi-linear $f(.)$

$$f(v) = v + v^3,$$

we have from (A1) the respective values of $D_\Sigma(\alpha)$ as $D_\Sigma(1) = 1.6$ and $D_\Sigma(3) = 1.1325$. Thus,

$$G(v_{in}) = 1.6 v_{in} + 1.1325 v_{in}^3, \tag{A2}$$

which for $v_{in} = 1$, gives our estimate for the input current as **$G(1) = 2.7325$**.

We now find the precise $i_{in}(1) = F(1)$ by determining $v_o$ from the "central" nodal equation

$$f(v_{in} - v_o) = f(v_o) + f(v_o/2) \tag{A3}$$

that for $f(v) = v + v^3$ and $v_{in} = 1$ takes the form

$$(1 - v_o) + (1 - v_o)^3 = v_o + v_o^3 + v_o/2 + (v_o/2)^3,$$

or

$$17 v_o^3 - 24 v_o^2 + 44 v_o - 16 = 0.$$

This equation has the positive root $v_o = 0.4350635$.

Substituting this value of $v_o$ into the input nodal equation,

$$F(v_{in}) = f(v_{in}) + f(v_{in} - v_o),$$

which for $v_{in} = 1$ is

$$F(1) = f(1) + f(1 - v_o)$$

$$= 1 + 1^3 + (1 - v_o) + (1 - v_o)^3, \tag{A4}$$

we obtain $F(1) = 2.7452378$. $G(1)$ is smaller than this correct value only by some **0.46 %**. In the examples of [4] too, $G(.)$ approximates $F(.)$ with the typical error being in the range from -0.4% to -0.8%.

### *A.1 Why do we consider the error of 0.46% of the analytical superposition to be small?*

That for $v_{in} = 1$ the circuit is strongly nonlinear is seen from (12); the ratio of the nonlinear term in $F(1)$ to the linear one is significant: $[1^3+(1-0.4350635)^3] / [1+(1-0.4350635)] = 1.1803/1.5649 = 0.754$. This relative value is much larger than the relative value 0.0046 of the imprecision in the analytical superposition.

It is also important to check that the circuit is not close to the specific examples of Section 4. Let us compare the high precision of the approximation of $F(1)$ by $G(1)$ with the relatively poor "constancy" of the nodal voltage $v_o(\alpha)$. The formula for $v_o(\alpha)$ found in [1] gives for $v_{in} = 1$: $v_o(1) = 0.4$ and $v_o(3) = 0.489583$. The relative difference between the above-found value $v_o = 0.435$ and $v_o(1)$, or $v_o(3)$, is of the order of 10%, which is much larger than the 0.46 % of the precision of the analytical superposition.

### *A.2 The linear and nonlinear parts of the F(.)*

As an illustration of the general fact, proved in Statement 1, let us observe for the above circuit that as $v_{in} \to 0$, $F(v_{in}) \sim D_{\underline{x}}(1)v_{in} = \varphi(1)v_{in}$ i.e. the first (linear) terms of the power series for $F(.)$ and $G(.)$ coincide.

In order to show this, we now write (A3) for arbitrary $v_{in}$:

$$(v_{in} - v_o) + (v_{in} - v_o)^3 = v_o + v_o^3 + v_o/2 + (v_o/2)^3, \quad (A5)$$

It is obvious that for $v_{in} = 0$, $v_o = 0$, and that as $v_{in} \to 0$, $v_o \to 0$, with the continuity. Since $v_o = O(v_{in})$, as $v_{in} \to 0$, the third-degree terms in (A5) become negligible with respect to the first-degree terms. Thus, as $v_{in} \to 0$ (A5) becomes the nodal equation of the asymptotically obtained *linear circuit of the same structure*:

$$v_{in} - v_o = v_o + v_o/2, \quad (A6)$$

from which $v_o = (2/5)v_{in}$. Substituting this value of $v_o$ into the input equation $F(v_{in}) = f(v_{in}) + f(v_{in} - v_o)$, *linearized for the same reasons*, we have:

$$F(v_{in}) = v_{in} + (v_{in} - v_o) = v_{in} + (v_{in} - (2/5)v_{in}), = (8/5)v_{in} = 1.6\, v_{in}$$

which indeed precisely equals the found *linear term* in $G(v_{in})$ given by (A2).

### *A.3 The basic inequality for $v_o$*

It is important for the proofs of [2] to observe that in the *f*-connection the obtained value of $v_o$ is *intermediate* with respect to the values that this nodal potential has in the composing circuits, the one with $\alpha = 1$, and the other with $\alpha = 3$.

Considering first $d_o = v_o/v_{in}$ for the $\alpha$-realization, we find by taking only linear terms in (A5), from this equation, that $d_o(\alpha = 1) = 2/5$, and by taking only cubic terms in (A5) that $d_o(\alpha=3) = 2(2+9^{1/3})^{-1} > d_o(\alpha=1)$. The latter inequality is natural since by increasing $\alpha$ we make the elements closer to hardlimiters, and (see the circuit) the right-hand branch becomes less relevant to the voltage division. Observe that $2(2+9^{1/3})^{-1}$ is already very close to ½.

Let us now use that in the *f*-connection (contrary to (4) related to $\alpha$-realization) $d_o$ depends on $v_{in}$, which is clear from (A5), and that as $v_{in}$ is increased from 0 to ∞, $d_o$ is increased from $d_o(\alpha = 1)$ to $d_o(\alpha = 3)$. The latter is because initially the linear, and, finally, the cubic, terms become dominant in $f(.)$ and in (A5). It needs to be proved, however, that with the changes in $v_{in}$, $d_o(v_{in})$ does not leave the interval $(d_o(\alpha=1), d_o(\alpha=3))$. For this it is sufficient to prove that $d_o(v_{in})$ is increased monotonically when $v_{in}$ is increased.

For the proof, we divide the identity (A5) by $v_{in}^3$, obtaining

$$\frac{1}{v_{in}^2}(1-d_o) = \frac{3}{2}\frac{d_o}{v_{in}^2} + \frac{9}{8}d_o^3 - (1-d_o)^3 \ . \tag{A7}$$

Differentiating the latter identity by $v_{in}$ we obtain

$$\frac{5d_o - 2}{v_{in}^3} = [\frac{5}{2}\frac{1}{v_{in}^2} + \frac{27}{8}d_o^2 + 3(v_{in} - d_o)^2]\frac{d}{dv_{in}}d_o . \tag{A8}$$

It is obvious from (A8) that $d(d_o)/dv_{in} > 0$ is provided by $5d_o - 2 > 0$, i.e. by $d_o > 2/5$. Thus, if we increase $v_{in}$ starting from the zero value (when $d_o = 2/5$), and obtain for the infinitesimal $v_{in}$, that $d_o > 2/5$, then $d_o(v_{in})$ necessarily continuously increases with the increase in $v_{in}$, and $d(d_o)/dv_{in} > 0$ all the time.

This situation indeed takes place. To show this, let us substitute into (A7), rewritten as

$$1 - d_o = \frac{3}{2}d_o + v_{in}^2 [\frac{9}{8}d_o^3 - (1-d_o)^3] \ , \tag{A7a}$$

$$d_o = \frac{2}{5} + \varepsilon \ , \tag{A9}$$

and assume the smallness of $\varepsilon$ to be of the order $v_{in}^2$, i.e. ignore in the derivations such terms as $\varepsilon v_{in}^2$, etc., which all are much smaller, as $v_{in} \to 0$. Substitution of (A9) into (A7a) gives then, with the asymptotic precision:

$$\varepsilon = -\frac{2}{5}[\frac{9}{8}(\frac{2}{5})^3 - (\frac{3}{5})^3]v_{in}^2 = 0.0576 v_{in}^2 > 0.$$

The positiveness of $\varepsilon$ proves, according to the above argumentation, that *for any* $v_{in}$, $d_o$ of the *f*-connection belongs to the interval $(d_o(\alpha=1), d_o(\alpha=3))$, i.e. it is some intermediate value, compared to the respective values on the separately taken composing circuits.

Quite similarly, using the dependence of $d_o$ on $v_{in}$ in the connected state, one can prove that for this topology, and $f(.) = D_m(.)^m + D_n(.)^n$, $m \neq n$, $d_o$ belongs to the interval $(d_o(\alpha=\min\{m,n\}), d_o(\alpha=\max\{m,n\}))$. For the *f*-connection, as $v_{in} \to 0$, and $v_{in} \to \infty$, the generalized $f(.)$ becomes only one of the two power-law terms, and along the line of the above argument, (A8) may be rewritten for this more general case as

$$\frac{d}{dv_{in}}d_o = \zeta(d_o, v_{in})[d_o - d_o(\alpha = \min\{m,n\})]$$

with an essentially positive function $\zeta(.,.)$ tending to zero as $v_{in} \to \infty$.

The intermediate nature of the value of the nodal potentials is typical in the *f*-connections, and, -- as [2] explains, -- the high precision of the analytical superposition observed in many examples is due to this circuit feature.

## **Appendix B: The mesh-currents (resistive, $v = f(i)$) formulation of the $\alpha$-circuit (or the '$\alpha$-test')**

We used a conductivity characteristic for representation of the $\alpha$-circuit and an input voltage source. However, a similar theory is easily developed if one uses a *resistive* characteristic and an input current source. Then, in particular, for the degree (power) of the power-low characteristic tending to infinity, we obtain each element of the associated "'$\alpha$-circuits" not as a voltage, but a current hardlimiter.

Using now the individual *resistive* characteristic (invert the axes of the graph of $i = f(v)$) and input *current source*, $i_{in}$, we shall apply, for simplicity of writing, the same notations, '*f*', *F*, '*D*', $\alpha$ and $\varphi$ as in the nodal-voltage formulation of the test, i.e. we write

$$v = f(i) = D\, i^\alpha, \qquad (B1)$$

and

$$v_{in} = F(i_{in}) = D\varphi(\alpha)\, i_{in}^{\,\alpha}.$$

The use of the same notations for the actually converted (regarding the units) functions, should not cause any problem, since one always uses either complete nodal-voltages or mesh-currents, description of a circuit. Only when comparatively considering in one equation (eq. (B3) below) $\varphi(.)$ for the resistive and conductive formulations, we shall use respective notations $\varphi(\alpha)_{meshes}$ and $\varphi(\alpha)_{nodes}$.

We can obtain the new $f(.)$ and $F(.)$ (or $\varphi(.)$) from the previous functions, i.e. we can first solve a circuit in the nodal voltage formulation, and then transfer to the mesh currents formulation. From the equation related to nodal voltages' description,

$$i = f(v) = D\, v^\alpha,$$

we have

$$v = D^{-1/\alpha}\, i^{-1/\alpha},$$

and making the transfer $\alpha \to 1/\alpha$, obtain the individual characteristic of the needed type

$$v = D^{-\alpha} i^{\alpha}.$$

Similarly, the equation

$$i_{in} = F(v_{in}) = D\varphi(\alpha) v_{in}^{\alpha}$$

yields the input characteristic of the needed form:

$$v_{in} = D^{-\alpha}[\varphi(1/\alpha)]^{-\alpha} i_{in}^{\alpha}. \tag{B2}$$

*Thus, in order to convert the nodal-voltages formulation to the mesh-currents formulation, we have first to replace $\alpha$ by $1/\alpha$, and then $D$ by $D^{-\alpha}$ and $\varphi(\alpha)$ by $[\varphi(1/\alpha)]^{-\alpha}$.*

The formula

$$\varphi_{meshes}(\alpha) = \frac{1}{[\varphi_{nodes}(1/\alpha)]^{\alpha}} \tag{B3}$$

is a generalization, for $\alpha \neq 1$, of the linear relation $r_{in} = 1/g_{in}$. The physically clear facts that for any circuit having, among other elements, the element a-b included, for any in $\alpha$  $\varphi(\alpha)_{nodes} >1$ and $\varphi(\alpha)_{meshes} <1$, agrees with the inverse-type relation between $\varphi_{meshes}$ and $\varphi_{nodes}$ given by (B3).

As $\alpha \to \infty$, we obtain now not voltage, but *current* hardlimiters. These hardlimiters could be obtained by inversion, using $f(.)$ from the nodal-voltages formulation, as $\alpha \to 0$.

As a simple example of the mesh-current formulation, consider the circuit shown in Fig. B1, which is of the same topology as (not "dual to") the circuit of Fig. A1 of Appendix A, but with *resistive* characteristic $f(.) = (.)^{\alpha}$, and with a current input source. The mesh currents $i_1$ and $i_2$ are now the unknowns.

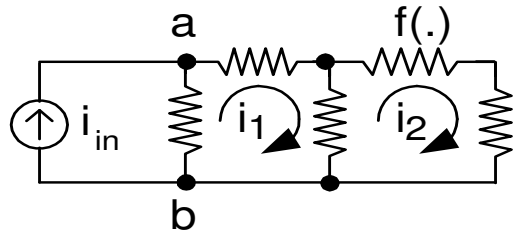

Fig. B1: The same topology as in Fig. A1 (Appendix A), but the elements have now *resistive* characteristic $v_s = f(i_s) = i_s^{\alpha}$, and at the input there is a *current* source.

Using (B3) one obtains the resistive $F(.)$ from the conductive one, as

$$\varphi(\alpha) = \frac{2+(1+2^{1/\alpha})^{\alpha}}{\left[1+2^{1/\alpha}+[2+(1+2^{1/\alpha})^{\alpha}]^{1/\alpha}\right]^{\alpha}}.$$

The same result is, of course, also easily found by direct calculation, in terms of mesh currents. The mesh equations (KVL) are:

$$-(i_{in} - i_1)^\alpha + i_1^\alpha + (i_1 - i_2)^\alpha = 0 \qquad (B4)$$

$$-(i_1 - i_2)^\alpha + 2i_2^\alpha = 0. \qquad (B5)$$

From (B5)

$$i_1 - i_2 = 2^{1/\alpha} i_2,$$

i.e.

$$i_2 = \frac{i_1}{1 + 2^{1/\alpha}},$$

and substituting this into (B4), we obtain

$$i_1 = \frac{i_{in}}{1 + [1 + \frac{2}{(1+2^{1/\alpha})^\alpha}]^{1/\alpha}}.$$

Since $v_{in} = (i_{in} - i_1)^\alpha$, we have

$$\varphi(\alpha) = \frac{(i_{in} - i_1)^\alpha}{(i_{in})^\alpha} = \left[1 - \frac{1}{1 + [1 + \frac{2}{(1+2^{1/\alpha})^\alpha}]^{1/\alpha}}\right]^\alpha \qquad (B6)$$

$$= \frac{2 + (1 + 2^{1/\alpha})^\alpha}{\left[1 + 2^{1/\alpha} + [2 + (1+2^{1/\alpha})^\alpha]^{1/\alpha}\right]^\alpha}.$$

## Acknowledgements

For the development of Appendix B, I am obliged to a comment by Prof. Bertram E. Shi, made at ECCTD'01, that such a version of the analytical superposition could be relevant to some applications of vision chips ([14,15] to an introduction to this topic), because current, and not voltage hardlimiters, arriving here as $\alpha \to \infty$, are widely used there. A tutorial, parallel to [3], given at ECCTD'01, had been initiated by Prof. Martti Valtonen.